\documentclass[12pt,a4paper,tightenlines,nofootinbib]{revtex4}
\usepackage{amsmath}
\usepackage{amssymb}
\usepackage{epsf}

\begin{document}

\author{Ariel M\'egevand} %
\altaffiliation{Member of CONICET, Argentina} %
\email[]{megevand@mdp.edu.ar} %
\affiliation{Departamento de F\'{\i}sica, Facultad de Ciencias
Exactas y Naturales, Universidad Nacional de Mar del Plata, De\'an
Funes 3350, (7600) Mar del Plata, Argentina}
\title{Gravitational waves from deflagration bubbles in first-order
phase transitions}

\begin{abstract}
The walls of bubbles in a first-order phase transition can propagate
either as detonations, with a velocity larger than the speed of
sound, or deflagrations, which are subsonic. We calculate the
gravitational radiation that is produced by turbulence during a phase
transition which develops via deflagration bubbles. We take into
account the fact that a deflagration wall is preceded by a shock
front which distributes the latent heat throughout space and
influences other bubbles. We show that turbulence can induce peak
values of $\Omega_{GW}$ as high as $\sim 10^{-9}$. We discuss the
possibility of detecting at LISA gravitational waves produced in the
electroweak phase transition with wall velocities $v_w\lesssim
10^{-1}$, which favor electroweak baryogenesis.
\end{abstract}

\maketitle

\section{Introduction} \label{intro}

One of the possible outcomes of a cosmological phase transition is
gravitational radiation. Although it is difficult to detect,
gravitational radiation provides a direct probe of the phase
transition dynamics, since it propagates freely until the present
epoch. Studies of the gravitational waves due to phase transitions at
the electroweak scale and beyond \cite{ewlisa} show that the signal
may be within the sensitivity range of LISA and the second generation
of space-based interferometers.

In a first-order phase transition, bubbles of the stable phase
nucleate and expand, converting the high-temperature phase into the
low-temperature one. Gravitational waves (GWs) are generated either
by the collisions of bubbles \cite{coll,kkt94,hk08,cds07} or by the
turbulence that is produced in the plasma due to the motion of bubble
walls \cite{hk08,cds07,kkt94,kmk02,dgn02,cd06,gkk07,kgr08}. As
bubbles expand, latent heat is released at the phase boundary. Part
of this energy raises the temperature of the plasma, and another part
is converted to bulk motions of the fluid. For phase transitions in
the early Universe, the Reynolds number is large enough for the wall
motion to produce turbulence. In general, the amplitude of GWs coming
from turbulence dominates over bubble collisions.

The bubble wall can propagate via two modes, namely, detonation and
deflagration. For a detonation, the phase transition front moves
faster than the speed of sound $c_{s},$ whereas a deflagration front
is subsonic. GWs produced by detonations have been extensively
investigated \cite{ewlisa,coll,hk08,cds07,kkt94,kmk02,dgn02,cd06}. In
this case, the bubble wall velocity $v_{w}$ depends only on the ratio
of the latent heat to the total energy density, $\alpha$ \cite{s82}.
Since $v_{w}>c_{s},$ no signal precedes the detonation front. Hence,
the dynamics of a wall is not influenced by other bubbles walls
(except in the collision regions). Furthermore, the kinetic energy
injected into the fluid is concentrated in a thin region behind the
wall. This simplifies the integration of the kinetic-energy density
profile. Besides, the temperature in the supercooled phase outside
the bubbles decreases due to the expansion of the Universe. As a
consequence, the nucleation rate $\Gamma $ increases exponentially.
This justifies modeling it by $\Gamma =\Gamma _{0}e^{\beta t}$, which
corresponds to linearizing the time dependence of the exponent.
Hence, $\beta ^{-1}$ is the only time scale in the problem and
determines the duration of the phase transition, $\Delta t\sim \beta
^{-1},$ and the mean bubble separation, $d\sim v_{w}\beta ^{-1}.$
Thus, the final result depends only on the parameters $\alpha$ and
$\beta$. These features simplify the calculation of GWs.

The case of deflagrations is more difficult. In contrast to
detonations, the deflagration front velocity depends on the viscosity
of the plasma and on the pressure difference between phases, which in
turn depends on the amount of supercooling. The wall propagates at a
subsonic velocity and is preceded by a supersonic shock front which
affects other bubbles. The shock wave distributes the latent heat
injected by the wall throughout space, causing a reheating of the
Universe and bulk motions of the fluid far away from the wall. For
these reasons, the case of deflagrations has not received as much
attention as the detonation case. In Ref. \cite{cds07} an analytic
expression is derived for the signal from bubble collisions. However,
in that work the fluid velocity is left as a free parameter.
Deflagrations have also been considered numerically in Ref.
\cite{kkt94}.

The spectrum of GWs for the deflagration case thus depends on
several parameters, which are difficult to estimate in a specific
model. As a consequence, the results for detonations have often been
used to investigate the GW production in different phase transitions
(see e.g. \cite{ewlisa}), disregarding the fact that the phase
transition may actually proceed via deflagration bubbles. Notice,
indeed, that this is likely the general case. For instance, for the
electroweak phase transition, estimations of the wall velocity give
in general subsonic values $v_{w}\sim 10^{-2}-10^{-1}$ \cite{vwew},
which are favorable for electroweak baryogenesis \cite{ewb}.

The aim of this work is to address the case of deflagration bubbles
with a realistic approach. Thus, we take into account the fact that
the shock waves coming from a bubble wall influence other bubbles. In
section \ref{dynamics} we describe the dynamics of deflagration
bubbles. One important feature is that, as a consequence of
reheating, the nucleation rate turns-off and the bubble wall
decreases during bubble expansion. Therefore, the approximation
$\Gamma\propto e^{\beta t}$ cannot be used in this case. Furthermore,
the smaller velocities at the collision time make bubble collisions
to be suppressed relative to turbulence as a source of GWs. In
section \ref{turb} we consider the velocity spectrum of the turbulent
fluid that arises from the motion of deflagration walls, and in
section \ref{gw} we follow the approach of Ref. \cite{cd06} to derive
the gravitational radiation produced by turbulence in the
deflagration case. In section \ref{result}  we estimate the
dependence of the amplitude and peak frequency on the latent heat,
the bubble size, and the wall velocity. We show that for a strong
enough electroweak phase transition, GWs from deflagration bubbles
may be detected at LISA. Our conclusions are summarized in section
\ref{conclu}.

\section{Phase transition dynamics\label{dynamics}}

A phase transition of the Universe occurs when the free energy of a
system depends on an order parameter $\phi $ (in general a Higgs
field), which develops a non-zero value. For a first-order phase
transition, there is a temperature range in which the free energy
$\mathcal{F}\left( \phi ,T\right) $ has two minima separated by a
barrier. Generally, the high-temperature minimum is $ \phi =0$,
corresponding to the symmetric phase, and the low-temperature one is
$\phi _{m}\left( T\right) \neq 0$, corresponding to the broken
symmetry
phase. At the critical temperature $T_{c}$, the two minima $\phi =0$ and $%
\phi =\phi _{m}$ have the same free energy. Below this temperature, $\phi
_{m}\left( T\right) $ becomes the global minimum. At a certain temperature $%
T=T_{0}$, the barrier disappears and $\phi =0$ becomes a maximum of the free
energy.

The energy density $\rho$ and entropy density $s$ are related to the
free energy density by  $\rho =Ts+\mathcal{F}$ and
$s=-d\mathcal{F}/dT$. We assume that the phase with $ \phi =0$ is
composed of radiation and false vacuum, i.e., $\mathcal{F}\left(
0,T\right) \equiv \mathcal{F}_{+}\left( T\right) =\rho _{\Lambda
}-\rho _{R}/3,$ where $\rho _{\Lambda }$ is the zero-temperature
energy density of the false vacuum, $\rho _{\Lambda}=\mathcal{F}(\phi
=0,T=0)=V\left( \phi =0\right) $, with $V(\phi)$ the zero-temperature
effective potential, and $ \rho _{R}$ is the energy density of
radiation, $\rho _{R}=g_{\ast }\pi ^{2}T^{4}/30,$ where $g_{\ast }$
is the number of relativistic degrees of freedom. Thus, for the
high-temperature phase we have $\rho _{+}=\rho _{\Lambda }+\rho
_{R}$, and for the low-temperature phase we have $\rho
_{-}=-T\mathcal{F}_{-}^{\prime }+\mathcal{F}_{-}$. The pressure is
given by $ p=-\mathcal{F}$. Hence, at $T=T_{c}$ both phases have the
same pressure. The latent heat $l$ is defined as the energy density
discontinuity $l=\Delta \rho =\rho _{+}(T_c)-\rho _{-}(T_c)$. Thus,
$l=T_{c}\Delta s=-T_{c}\Delta \mathcal{F}^{\prime }$. We note that in
previous works on GW generation the latent heat is sometimes confused
with the false vacuum energy density (or even with the free energy
density). In fact, notice that in general $l$ can be anywhere in the
range $0<l<\rho _{\Lambda }+\rho _{R}(T_{c})$. On the other hand, it
can be shown that $\rho _{\Lambda }$ is bounded by $0<\rho _{\Lambda
}<\rho _{R}(T_{c})/3$ (assuming that the vacuum energy vanishes at
$T=0$) \cite{ms08}.

The nucleation and growth of bubbles has been extensively
investigated (see e.g.
\cite{s82,vwew,ewb,ms08,bubbles,nucl,h95,m04,m06,ma05,ikkl94}).
According to the standard picture, bubbles of the stable phase
nucleate with a rate \cite{nucl}
\begin{equation}
\Gamma \approx T^{4}e^{-S_{3}/T}, \label{gamma}
\end{equation}%
where $S_{3}\left( T\right) $ is the three-dimensional instanton
action, which coincides with the free energy of a critical bubble.
The nucleation rate is extremely sensitive to the temperature in the
range $T_{0}<T<T_{c}.$ At the critical temperature the radius
$R_{c}$ of the critical bubble becomes infinite, so $S_{3}=\infty $
and $\Gamma =0.$ In contrast, at $ T=T_{0}$ the radius vanishes, so
$S_{3}=0$ and $\Gamma \sim T_{c}^{4}$, which is an extremely large
rate in comparison to $H^{4}\sim (T_{c}^{2}/M_{P})^{4}$. Thus,
bubbles begin to nucleate at an intermediate temperature $T_{\Gamma
}$ for which $\Gamma \approx H^{4}.$ To get an idea of the
dependence of $\Gamma$ on $T$, consider the thin-wall approximation,
which is valid near the critical temperature. In this case we can
write $S_{3}=-\Delta \mathcal{F} 4\pi R_{c}^{3}/3+\sigma 4\pi
R_{c}^{2},$ where $\sigma $ is the surface tension of the bubble
wall. Then we obtain the critical radius $ R_{c}=2\sigma /\Delta
\mathcal{F}$ and the action $S_{3}=16\pi \sigma ^{3}/3(\Delta
\mathcal{F})^{2}.$ For $T\approx T_{c}$ we can also approximate the
free energy difference by $\Delta \mathcal{F}\approx
l(T_{c}-T)/T_{c}.$ Thus, we obtain
\begin{equation}
\frac{S_{3}(T)}{T}\approx \frac{16\pi \sigma ^{3}T_{c}}{3l^{2}
(T_{c}-T)^{2}}.
\label{s3}
\end{equation}%
Once a bubble is nucleated, it begins to grow. The bubble radius
rapidly becomes much larger than $R_c$. Due to the viscosity of the
plasma, the bubble walls immediately reach a terminal velocity
$v_{w}$ which is determined by the pressure difference $\Delta
p=-\Delta \mathcal{F}$ and the friction with the surrounding
particles. It can be approximated by $v_{w}(T)=\Delta
\mathcal{F}(T)/\eta ,$ where $\eta $ is a friction coefficient.
Using again the linear approximation for $\Delta \mathcal{F}$ we
obtain
\begin{equation}
v_{w}\approx l(T_{c}-T)/\eta T_{c}.  \label{vw}
\end{equation}%
For strong phase transitions, the latent heat $l$ and the amount of
supercooling $(T_{c}-T_{\Gamma})/T_c$ will be considerable. Notice,
however, that the velocity depends also on the viscosity and can be
subsonic even in this case.

We shall assume that the wall propagates as a deflagration front.
Therefore, the wall velocity is lower than the speed of sound in the
relativistic plasma, $ c_{s}=\sqrt{1/3}$. In this case, a shock front
precedes the wall with a velocity $v_{sh}\gtrsim c_{s}$.
Consequently, the shock front of a bubble soon influences other
bubbles. For $v_{w}\ll c_{s}$, the latent heat is transmitted away
from the wall and is quickly distributed throughout space. This
effect can be taken into account by considering a homogeneous
reheating of the plasma as latent heat is being injected \cite{h95}.
Even with this simple approximation, the evolution of the phase
transition must be computed numerically. Therefore, it is not
straightforward to find relations between thermodynamic parameters
such as $l$ or $\sigma$ and the quantities that characterize the
dynamics, such as the duration of the phase transition or the bubble
number density.

Nevertheless, as we shall see, the general features of the dynamics
of slow bubble walls provide a good deal of information for the
computation of gravitational waves. As a consequence of reheating,
the free energy difference $\Delta \mathcal{F}(T)$ decreases and,
according to Eq. (\ref{vw}), the bubble expansion slows down. This
effect will be important if the latent heat is comparable to the
difference $\delta \rho=\rho _{R}\left( T_{c}\right) -\rho _{R}\left(
T_{\Gamma }\right) $. Thus, if $l\gg \delta\rho ,$ the temperature
will get very close to the critical one\footnote{This is quite
generally the case. For strong phase transitions, we have
$l\sim\rho\gg\delta\rho$. For weaker phase transitions, the latent
heat is smaller, but so is $\delta\rho$.}. Then, the velocity will
decrease significantly and a long phase coexistence stage will take
place before the transition completes \cite{m04,ms08,m06}. In any
case, due to the exponential dependence of the nucleation rate on
temperature, bubble nucleation turns off as soon as temperature
begins to raise \cite{m04,ma05}. Indeed, notice that the temperature
$T(t)$ has a minimum, which separates the supercooling and reheating
stages. According to Eqs. (\ref{gamma}) and (\ref{s3}), the
nucleation rate has a maximum at that time, and due to the extreme
dependence of $\Gamma$ on $T_c-T$, the maximum must be a sharp peak.
Hence, the minimum temperature gives the nucleation temperature
$T_{\Gamma}$. Most bubbles are created in a very short time interval
$\delta t_{\Gamma }$ around the time $t_{\Gamma }$ corresponding to
this temperature.

For $t>t_{\Gamma}$, the number density of bubbles $n_b$ remains
constant. The final size of bubbles is given by $d\sim n_b^{-1/3}$.
Most bubbles begin to expand at $t\approx t_{\Gamma}$ with velocity
$v_w\equiv v_i$. During reheating the wall velocity decreases, and by
the time bubbles percolate $v_w$ will be in general much smaller than
the initial velocity $ v_{i}$. As a consequence, the GW signal from
bubble collisions will be  too low. Therefore, we will consider only
GWs from turbulence. The main turbulence will be generated during the
reheating stage, when $v_{w}$ is still close to $v_{i}.$ The
frequency and amplitude of the GWs will depend on the dynamics of the
phase transition, which is involved. However, as we shall see, the
result will be essentially determined by a few parameters.

\section{Turbulence from deflagration bubbles\label{turb}}

Turbulence from stirring in cosmological  phase transitions has been
extensively studied \cite{kkt94,kmk02,dgn02,cd06,gkk07,kgr08}. When
turbulence is fully developed, a cascade of energy is established
from larger to smaller length scales, as eddies of each size break
into smaller ones. We define the energy dissipation rate per unit
enthalpy for a given
momentum scale $k=2\pi /L,$%
\begin{equation}
\varepsilon _{k}\equiv \frac{1}{w}\left.
\frac{d\rho _{\mathrm{turb}}}{dt}\right\vert_{\rm in},
\label{epsk0}
\end{equation}%
where $w=\rho +p$ is the enthalpy density, $\rho $ is the total
energy density of the fluid, $p$ is the pressure, and $ \rho
_{\mathrm{turb}}=w\langle v^{2}\rangle /2$ is the kinetic energy
density of turbulence. Since the phase transition occurs in the
radiation dominated epoch, we will make for simplicity the usual
assumption\footnote{%
In fact, the energy density at the beginning of the phase transition
is $\rho _{+}=\rho _{R}+\rho _{\Lambda }$. For a thermal phase
transition the false-vacuum energy is bounded by $\rho _{\Lambda
}<\rho _{R}/3$ and is in general $ \rho _{\Lambda }\ll \rho _{R}$
\cite{ms08}. Part of $\rho_+$ is liberated as latent heat, so that at
the end of the transition we have $\rho _{-}=\rho _{+}-l.$ } $\rho
\approx \rho _{R}$. Hence, $p\approx p_R=\rho_R/3$ and $w\approx
w_R=4/3\rho_R$. If the process is stationary, $\varepsilon _{k}$
gives the rate at which turbulent energy is received at the scale $k$
from higher length scales (and transferred to smaller length scales).
If the external source stirs the fluid at a single scale $k_{S}=2\pi
/L_{S}$, we have a constant rate $ \varepsilon _{k}\equiv \varepsilon
$ for scales $L<L_{S}$ \cite{dgn02}. In this case the turbulent
energy in the cascade is characterized by the Kolmogoroff spectrum
\begin{equation}
E(k)\equiv \frac{1}{w}\frac{d\rho
_{\mathrm{turb}}}{dk}\approx\varepsilon ^{2/3}k^{-5/3}.  \label{ek}
\end{equation}%
The cascade stops at the damping scale $k_D=2\pi/L_D$, with $%
L_{D}\ll L_{S},$ at which the fluid viscosity dissipates the injected energy
into heat.

As usual, we assume that the fluid is incompressible and
statistically isotropic and homogeneous. Then, the Fourier transform
of the velocity has the two point correlation function
\begin{equation}
\langle v_{i}(\mathbf{k})v_{j}^{\ast }(\mathbf{q})\rangle =
(2\pi )^{3}\delta ^{3}\left( \mathbf{k}-\mathbf{q}\right)
\left( \delta _{ij}-\mathbf{\hat{k}}_{i}\mathbf{\hat{k}}
_{j}\right)  P(k), \label{twopoint}
\end{equation}%
where the angular brackets mean a statistical average. Notice that
$\langle \mathbf{v}^{2}(\mathbf{x})\rangle $ gives the kinetic energy
density per unit enthalpy density of the fluid. Therefore, the
velocity spectrum $P(k)$ can be related to the energy spectrum
$E(k)$. The relation is $E(k)=k^{2}P(k)/2\pi ^{2}$ \cite{landau}.
From Eq. (\ref{ek}), we have
\begin{equation}
P(k)\approx \pi ^{2}\varepsilon ^{2/3}k^{-11/3}  \label{pk}
\end{equation}
in the inertial range $k_S\ll k\ll k_D$. The power spectrum $P(k)$
beyond this range was obtained in Ref. \cite{cd06} from an ansatz
for the real-space correlation function. We have
\begin{equation}
P(k)\approx 2\pi \langle v^2\rangle \left\{ \begin{array}{ll}
\frac{2}{765}L_{S}^{5}k^{2} & \text{for }k\ll k_{S} \\
\frac{55}{81}\sqrt{3}\Gamma \left( \frac{2}{3}\right)
L_{S}^{-2/3}k^{-11/3}
& \text{for }k_{S}\ll k\ll k_{D} \\
0 & \text{otherwise},%
\end{array}%
\right. \label{pkcd}
\end{equation}
and the normalization of the spectrum is given by
\begin{equation}
\langle v^2\rangle \approx\left(\varepsilon L_S\right)^{2/3},
\label{v2}
\end{equation}
which corresponds to the fluid velocity on the largest scale $L_S$
(see below).

For stationary turbulence, the energy dissipation rate in Eq.
(\ref{epsk0}) must equal the power that is injected by the source.
Thus, for the case of expanding bubbles we have
\begin{equation}
\varepsilon=\frac{1}{w}\kappa l \frac{df_{b}}{dt}, \label{drhoturb}
\end{equation}%
where  $f_{b}$ is the fraction of volume occupied by bubbles and
$\kappa $ is an efficiency factor which quantifies the fraction of
latent heat that goes into kinetic energy of the fluid (a fraction
$1-\kappa $ goes into thermal energy and causes reheating of the
plasma). The shock wave in front of the wall sets the fluid moving
outward with a velocity $v_{f}$. Hence, the injected kinetic-energy
density is
\begin{equation}
\rho _{\mathrm{kin}}=\frac{1}{2}wv_{f}^{2}.  \label{rhoturb}
\end{equation}%
As the phase transition front moves a distance  $v_w \delta t$, the
energy released is proportional to $l v_w \delta t$. Assuming for
simplicity a constant fluid velocity up to a distance $\sim c_s t$,
the kinetic energy injected in the time $\delta t$ will be
proportional to $\rho_{\rm kin}c_s\delta t$. Thus, we have
\begin{equation}
\kappa =\frac{\rho _{\mathrm{kin}}c_s}{lv_w}.  \label{defkappa}
\end{equation}%
We still need to determine the fluid velocity $v_{f}$ appearing in
Eq. (\ref{rhoturb}). The conservation of energy and momentum
$\partial _{\mu }T^{\mu \nu }=0$ can be applied to the wall
discontinuity to obtain relations for the quantities on both sides of
the wall. Assuming a stationary solution and non-relativistic
velocities, one obtains the equation (see e.g. \cite{landau,ikkl94})
\begin{equation}
w_{+}v_{+}=w_{-}v_{-},  \label{encons}
\end{equation}%
where $w$ is the enthalpy, $v$ is the velocity of the fluid in the
rest frame of the phase transition front, and the plus and minus
signs stand for the high- and low-temperature phase regions,
respectively. In the rest frame of the center of the bubble, the
fluid inside the bubble is at rest, the wall velocity is
$v_{w}=-v_{-}$, and the fluid velocity in front of the wall is
$v_{f}=v_{+}-v_{-}.$ Using Eq. (\ref{encons}) we obtain
\begin{equation}
v_{f}=\frac{w_{+}-w_{-}}{w_{+}}v_{w}\approx\frac{\Delta \rho +
\Delta p}{w_{R}}
v_{w},  \label{vfenth}
\end{equation}%
where $\Delta \rho $ and $\Delta p$ are the energy-density and
pressure differences across the wall. Notice that, even if the
temperature is homogeneous, $\rho \left( T\right) $ and $p\left(
T\right) $ are different in each phase. $\Delta \rho (T)$ and $\Delta
p(T)$ depend on the amount of supercooling. For simplicity, we will
assume that the temperature remains close to the critical one, so
that $\Delta p\approx 0$ and the energy density discontinuity is
given by the latent heat, $\Delta \rho \left( T_{c}\right) \equiv l.$
Therefore, we have
\begin{equation}
v_{f}\approx\left( l/w_{R}\right) v_{w}.  \label{vf}
\end{equation}%
From Eqs. (\ref{rhoturb}), (\ref{defkappa}), and (\ref{vf}) we obtain
\begin{equation}
\kappa =\frac{1}{2}(l/w_{R})v_{w}c_s
\end{equation}%
Thus, Eq. (\ref{drhoturb}) gives
\begin{equation}
\varepsilon =(\kappa l/w)\dot{f}_{b}=(1/2)v_{w}c_s\alpha ^{2}
\dot{f}_{b},  \label{eps}
\end{equation}%
where we have defined the ratio of the latent heat to the enthalpy
density, $\alpha \equiv l/w_{R}$.

For the validity of the Kolmogoroff spectrum, it is important that
the energy is injected at a single scale $L_S$. For the kind of
phase transitions we are interested in, all the bubbles are formed
in a short interval $\delta t_{\Gamma }$ around the time $t_{\Gamma
}. $ Hence, the number density of bubbles $n_{b}$ is set at
$t\approx t_{\Gamma}$. For $t>t_{\Gamma }$, the number of bubbles
remains constant, and the wall velocity $v_{w}\left(T( t)\right) $
is the same for all bubbles. The bubble radius is
$R(t)=\int_{t_{\Gamma }}^{t}v_{w}(t^{\prime })dt^{\prime },$ and the
distribution of sizes has a small width $\delta R\approx
v_{w}(t_{\Gamma })\delta t_{\Gamma }\ll R.$ At the moment of
collision the bubble size is given by the distance between bubble
centers, $d\sim n_{b}^{-1/3}.$

In fact, turbulence begins as soon as the {\em shock} fronts collide.
The first shocks are emitted at $ t\approx t_{\Gamma }$ and collide
when they reach the size $d,$ after a time $\sim d/c_{s}$. This time
is much smaller than the total duration $\Delta t$ of the phase
transition. Indeed, for deflagrations the wall velocity is $v_w<c_s$.
In addition, as explained in section \ref{dynamics}, $v_w$ decreases
significantly from its initial value $v_i$. Therefore, we have
$\Delta t\gg d/v_i>d/c_s$.  Moreover, as mentioned in section
\ref{intro}, in the general case the initial velocity may be already
$v_i\ll c_s$, as suggested by the electroweak case.  When shock
fronts collide, they lose their spherical symmetry. Bubbles continue
expanding, and their walls continue injecting energy into the fluid
until all space is filled up. Thus, turbulence begins very soon, at
$t\approx t_{\Gamma },$ and remains until the phase transition is
complete. As bubbles expand, the latent heat is taken away at the
speed of sound, always stirring the fluid at the same scale
$L_{S}\sim d.$ Hence, it is reasonable to assume a Kolmogoroff
spectrum with $k_{S}\sim 2\pi /d.$ Notice that in the detonation
case, in contrast, there will be much more small bubbles (with $L\ll
d$) than large ones (with $L\sim d$), due to the exponentially
increasing nucleation rate $\Gamma=\Gamma_0 e^{\beta t}$. Thus, the
small bubbles will in principle modify the Kolmogoroff spectrum in
the detonation case.

It is important to estimate the turnover time scale of an eddy,
$\tau _{L}\approx L/v_{L},$ where $v_{L}$ is the characteristic
fluid velocity on a length scale $L$. The rate at which an eddy
breaks into smaller ones is usually assumed to be roughly $\sim \tau
_{L}^{-1}.$ Thus, the rate at which energy is transferred in the
cascade is $\varepsilon \sim v_{L}^{2}\tau _{L}^{-1}.$ Therefore, we
have $v_{L}\sim (\varepsilon L)^{1/3}$ and $\tau _{L}\sim
\varepsilon ^{-1/3}L^{2/3}.$ The turnover frequency is given by
$\omega_k=\omega_S(k/k_S)^{2/3}$, with $\omega_S=v_{L_S}/L_S$.

Both $v_{L}$ and $\tau _{L}$ increase with $L$. The maximum length
scale is the characteristic scale of the source $L_{S}\sim d$. In
general, $d$ is well inside the horizon \cite{ms08,ma05,h95}. Hence,
the time scale for the establishment of a cascade, which is on the
order of the maximum turnover time $\tau _d,$ will be at most on the
order of the Hubble time. This justifies neglecting the expansion of
the Universe in the description of turbulence. On the other hand, it
is important to compare the characteristic turnover time with the
duration of the phase transition.  For the case of detonations, it
has been shown that $\tau _d$ is always larger than $\Delta t$ (see
e.g. \cite{kmk02,cd06}). This result is more general. Indeed, since
$\tau _d\sim d/v_d$, $\Delta t \sim d/v_w$, and the fluid velocity is
always smaller than the wall velocity,  we have $\tau_d>\Delta t$.
Since the stirring source lasts less than $\tau_d$, turbulence is not
stationary. Nevertheless, the cascade of energy develops, and it was
argued \cite{kmk02} that, for the generation of gravitational waves,
we can assume stationary turbulence with a duration $\sim
\tau_{L_S}=\tau_d$.

\section{Relic gravitational waves\label{gw}}

We aim to calculate the spectrum we would observe today for GWs
originated in a cosmological phase transition at time $t=t_*$ and
temperature $T=T_*$. As explained in Sec. \ref{dynamics}, in the
deflagration bubble scenario the wall velocity is in general
considerably smaller at the time of percolation than at the
beginning of bubble expansion. Therefore, we expect a weak signal
from bubble collisions. In view of that, we shall only consider the
source provided by turbulence. The calculation of the GW spectrum
from primordial turbulence has been improved in the last years
\cite{cd06,gkk07}. Here we will use the results of Ref. \cite{cd06}
(see the discussion in Sec. \ref{conclu}). In this section we
briefly review the derivation and write down the results in terms of
variables that are suitable for the deflagration bubble scenario.

For a stochastic background of gravitational waves, the spectrum is
characterized by the quantity \cite{maggiore00}
\begin{equation}
\Omega _{GW}\left( f\right) =\frac{1}{\rho _{c}}\frac{d\rho
_{GW}}{d\log f}, \label{omega}
\end{equation}%
where $\rho _{GW}$ is the energy density of the GWs, $f$ is the
frequency, and $\rho _{c} $ is the critical energy density today,
$\rho _{c}=3H_{0}^{2}/8\pi G.$ Thus, the gravitational wave energy
density per unit logarithmic frequency is defined by the relation
\begin{equation}%
\rho_{GW}=\int \frac{df}{f}\frac{d\rho _{GW}}{d\log f}.
\end{equation}
For a stochastic background, $d\rho _{GW}/d\log f$ is given by the
ensemble average of the Fourier amplitudes of the tensor metric
perturbation $h_{ij}$. The energy density of gravitational waves is
\begin{equation}%
\rho_{GW}(\mathbf{x},t)=\frac{\langle \partial_th_{ij}(\mathbf{x},t)
\partial_th_{ij}(\mathbf{x},t)\rangle}{16\pi G},
\end{equation}
where the brackets denote the ensemble average. The source of
$h_{ij}(\mathbf{x},t)$ is the transverse and traceless piece of the
stress-energy tensor, which for a turbulent plasma is given by $
T_{ij}\left( \mathbf{x},t\right) =wv_{i}\left( \mathbf{x},t\right)
v_{j}\left( \mathbf{x},t\right) . $ Thus, the source for tensor
perturbations on each mode $\mathbf{k}$ is the anisotropic stress
$\Pi_{ij}(\mathbf{k})$,
%=(P_{il}P_{jm}-1/2P_{ij}P_{lm})T_{lm}(\mathbf{k})$
which involves a convolution $\int
d^3qv_i(\mathbf{q},t)v_j(\mathbf{k}-\mathbf{q},t)$ (see \cite{cd06}
for details). Therefore, the energy density spectrum, which involves
the average $\langle \Pi_{ij}(\mathbf{k}) \Pi_{ij}^*(\mathbf{q})
\rangle$, can be related to the velocity spectrum $P(k)$ by means of
Eq. (\ref{twopoint}).

Notice that the velocity correlation function (\ref{twopoint}) does
not oscillate in time.  However, the gravitational radiation is
produced by the turbulent eddies, which have a turnover frequency
$\omega_k=v_L/L$, with $k=2\pi/L$. The oscillatory behavior of the
source, which is relevant for the generation of GWs, is lost in the
statistical average. One can account for the turnover frequency  by
replacing $v_i(\mathbf{k})\to v_i(\mathbf{k})e^{i\omega_k t}$ for
$k_S<k<k_D$. As a consequence, the source for $h_{ij}(\mathbf{k},t)$
can be modeled as $e^{i2\bar{\omega}
t}\Pi_{ij}(\mathbf{k})\Theta(t-t_{\rm in})\Theta(t_{\rm fin}-t)$,
where $\bar{\omega}=\omega_k$ for $k>k_S$, $\bar{\omega}=\omega_{S}$
for $k<k_S$, and the Heaviside functions limit the source to the
interval $t_{\rm in}<t<t_{\rm fin}$. We set $t_{\rm fin}-t_{\rm
in}=\tau_{L_S}$ and $t_{\rm fin}=t_*$. For $t> t_*$, the
gravitational wave propagates freely, with the dispersion relation
$\omega=k$. Hence, the frequency $f$ in Eq. (\ref{omega}) is given
by the wave number $k$ of the source. The amplitude of the wave is
proportional to $\Pi_{ij}(\mathbf{k})$ and depends on the frequency
$\bar{\omega}$ of the source.

The spectrum of GWs is  obtained from the quantity $\langle
\dot{h}_{ij}(\mathbf{k},t) \dot{h}_{ij}(\mathbf{q},t)\rangle$. It is
proportional to the dimensionless function
\begin{equation}
A(\bar{\omega},k)=\frac{\left|e^{i(2\bar{\omega}-k)/\omega_S}-
1\right|^2}{(k
-2\bar{\omega})^2L_S^2}+\frac{\left|e^{i(2\bar{\omega}+k)/\omega_S}-
1\right|^2}{(k +2\bar{\omega})^2L_S^2}
\end{equation}
and to $\langle \Pi_{ij}(\mathbf{k})\Pi_{ij}^*(\mathbf{q})\rangle$.
The latter is a four-point spectral function of the velocity, and
must be reduced in order to use the two-point function
(\ref{twopoint}). This is usually done by using Wick's theorem,
although the velocity field is not
Gaussian. We have %\cite{cd06}
\begin{equation}
\langle
\Pi_{ij}(\mathbf{k})\Pi_{ij}^*(\mathbf{q})\rangle=w^2
\delta(\mathbf{k}-\mathbf{q})
\int
d^3pP(\mathbf{p})P(|\mathbf{k}-\mathbf{p}|)(1+\gamma^2)(1+\beta^2),
\label{piij}
\end{equation}
where $\gamma=\hat{\mathbf{k}}\cdot\hat{\mathbf{p}}$, $\beta =
\hat{\mathbf{k}}\cdot\widehat{\mathbf{k}-\mathbf{p}}$. Analytical
approximations exist for the integral in Eq. (\ref{piij}) for large
and small scales \cite{kmk02,cd}. It can then be evaluated using
Eqs. (\ref{pkcd}) and (\ref{v2}).  Assuming radiation domination at
$T=T_{\ast },$ the enthalpy $w$ in Eq. (\ref{piij}) can be
approximated by $w_R=4\rho_R/3$, and  the Hubble rate is given by
$H_{\ast }^2=8\pi G\rho _{R}/3$. Finally,  at $t=t_*$ we
have\footnote{In Ref. \cite{cd06} the calculations were done using
conformal time $\eta$ and comoving variables
$\mathbf{k},\mathbf{x},L$, etc. Dimensionless combinations such as
$HL$ are readily translated to physical variables, since the scale
factor $a$ cancels out.}
\begin{equation}
\rho_{GW}(k,t_*)=\frac{\rho_R(t_*)}{4\pi}(H_*L_S)^2 (\varepsilon
L_S)^{4/3}\left\{
\begin{array}{ll}
\frac{2}{13}A(\omega_{S},k)(kL_S)^3 & \text{for }k < k_{S} \\
A(\omega_{k},k)(kL_S)^{-2/3}
& \text{for }k_{S}< k< k_{D} \\
0 & \text{otherwise}.%
\end{array}%
\right.
\end{equation}

The function $A(\bar{\omega},k)$ gives a different spectral
dependence whether the largest eddy velocity $v_{L_S}$ is below or
above $1/2$, but the GW spectrum always peaks at $k=k_S$. For
deflagrations the velocity of the fluid is smaller than the bubble
wall velocity, which is smaller than that of sound ($c_s\approx
0.58$). Thus, only in the limit in which both $v_{L_S}\approx v_w$
and $v_w\approx c_s$, we will have $v_{L_S}> 1/2$. We shall deal
with wall velocities $v_w\lesssim 0.1$. Therefore, we only consider
the case $v_{L_S}< 1/2$, for which the energy spectrum can be
approximated by
\begin{equation}
\rho_{GW}(k,t_*)=\frac{\rho_R(t_*)}{2\pi}(H_*L_S)^2 (\varepsilon
L_S)^{4/3}\left\{
\begin{array}{ll}
(k/k_S)^3/v_{L_S}^2 & \text{for }k<2v_{L_S}k_S \\
4(k/k_S)
& \text{for } 2v_{L_S}k_S< k < k_{S} \\
4(k/k_S)^{-8/3} & \text{for }k_{S}< k< k_{D} \\
0 & \text{otherwise}.%
\end{array}%
\right.
\label{omegaast}
\end{equation}

The GWs generated at time $t_{\ast }$ redshift due to the expansion
of the Universe. The energy density scales like $a^{-4}$, and the
frequency like $a^{-1}$. Therefore, the spectrum today is given by
Eq. (\ref{omegaast}), with $\rho_R(t_*)$ replaced with $\rho_R(t_0)$
and the wave numbers replaced with the corresponding frequencies.
Hence,
\begin{equation}
\Omega_{GW}(f,t_0)=\frac{\Omega_R(t_0)}{2\pi} \left(\frac{L_S}{H_*^{-1}}
\right)^{10/3}
\left(\frac{\varepsilon}{H_*} \right)^{4/3}\left\{
\begin{array}{ll}
(f/f_p)^3/v_{L_S}^2 & \text{for }f<2f_S \\
4(f/f_p)
& \text{for } 2f_S< f < f_p \\
4(f/f_p)^{-8/3} & \text{for }f_p< f< f_D \\
0 & \text{otherwise},%
\end{array}%
\right. \label{omeps}
\end{equation}
where $\Omega_R(t_0)=\rho_R/\rho_c\approx 4.6\times 10^{-5}$
\cite{pdg06}, and $f_S$, $f_p$, $f_D$ are the redshifted frequencies
  corresponding respectively to the
frequency of the largest eddies $f_{S\ast }=v_{L_S}L_S^{-1}$, the
peak frequency $f_{p\ast }=L_{S}^{-1}$, and the dissipation frequency
$f_{D\ast }=L_D^{-1}$. A frequency $f_*$ redshifted to today is given
by $f_0=f_*a_*/a_0$. The ratio of the scale factor at $t_{\ast }$ to
the scale factor today is
\begin{equation}
\frac{a_{\ast }}{a_{0}}\approx 8\times 10^{-16}\left( \frac{100}{g_{\ast }}%
\right) ^{1/3}\frac{100GeV}{T_{\ast }}.
\end{equation}%
It is useful to express $f_0$ in terms of $f_{\ast}/H_{\ast}$. The
Hubble rate $H_{\ast }=\sqrt{8\pi G\rho _{R}/3}$ can be written as
\begin{equation}
H_{\ast }\approx 2\times 10^{10}Hz\left( \frac{g_{\ast
}}{100}\right) ^{1/2}\left( \frac{T_{\ast }}{100GeV}\right) ^{2},
\end{equation}%
where we have used the relations $G=M_{Pl}^{-2},$ with
$M_{Pl}=1.22\times 10^{19}GeV$, and $1GeV\approx 1.5\times
10^{24}Hz.$ Therefore, the frequency  today is given by
\begin{equation}
f_{0}=1.6\times 10^{-5}Hz\frac{T_{\ast }}{100GeV}%
\left( \frac{g_{\ast }}{100}\right) ^{1/6}\frac{f_{\ast }}{H_{\ast }}.  \label{f0}
\end{equation}%

\section{Gravitational radiation from the phase transition\label{result}}

As we have seen, the scale $L_{S}$ is given by the bubble
separation. Therefore, the peak  frequency is
\begin{equation}
f_{p}=1.6\times 10^{-2}mHz\left( \frac{g_{\ast }}{100}\right) ^{1/6}
\frac{T_{\ast }}{100GeV}\frac{H_{\ast }^{-1}}{d}.  \label{fs}
\end{equation}%
The distance $d$ depends on the dynamics of the phase transition. It
can vary from values $d/H_{\ast }^{-1}\sim 10^{-5}$ for weakly
first-order phase transitions, to values $d/H_{\ast }^{-1}\sim
10^{-1}$ for strongly first-order phase transitions (see e.g.
\cite{ms08,ma05,h95}). For $T_{*}\gtrsim 100GeV$, we see that
millihertz frequencies (corresponding to the peak sensitivity of
LISA) are obtained for relatively large values of the bubble size,
$d/H_{\ast }^{-1}\gtrsim 10^{-2}$ (for $g_*\sim 100$).

Setting $L_{S}=d$ in Eq. (\ref{omeps}) we obtain, for the maximum of
the spectrum today,
\begin{equation}
\left.\Omega _{GW}\right\vert _{\rm peak}=3\times
10^{-5}\left(\frac{d}{H_*^{-1}}\right)^{10/3}
\left(\frac{\varepsilon}{H_*} \right)^{4/3}, \label{om}
\end{equation}%
According to Eq. (\ref{eps}), the rate $\varepsilon$ at which energy
is injected into the fluid is proportional to $ v_w\dot{f}_b$. The
speed of bubble expansion may vary considerably due to reheating
during the phase transition.  Nevertheless, we have seen that the
turbulence which sources the gravitational radiation lasts longer
than  the phase transition. Hence, the generation of GWs is not
affected by the details of the time dependence of $\varepsilon$, and
we can use in Eq. (\ref{om}) a mean value $\bar{\varepsilon}$ which
involves a time average of $v_w(t)\dot{f}_b(t)$. A precise
calculation of the parameters $d$ and $\bar{\varepsilon}$ requires a
numerical computation of the phase transition. We will address such
computation elsewhere \cite{progress}. Below, we find an
approximation for $\dot{f}_b$ as a function of $v_w$ and $d$. The
bubble separation $d$ also depends on the wall velocity. The larger
the initial velocity $v_i$, the quicker the reheating and the sooner
the turn-off of the nucleation rate. Roughly, we have $d\sim
n_b^{-1/3}\propto v_i$ \cite{m04}. Unfortunately, the number density
of bubbles $n_b$ is extremely sensitive to the dynamics of the phase
transition. This prevents any sensible analytical approximation for
$d$ as a function of the parameters of the model.

The bubble expansion rate depends on the wall velocity and the bubble
size. Roughly, $\dot{f}_{b}\sim Fd^{2}\bar{v}_{w}/d^{3}=F
\bar{v}_{w}/d,$ where $F$ is a geometrical factor (for a spherical
bubble, $F=4\pi $), and  $\bar{v}_{w}$ is the average wall velocity.
In general, it will not be a good approximation to take a plain time
average over the entire duration of the phase transition. The bubble
expansion will slow down significantly due to reheating. Therefore,
the turbulence that is generated while the wall velocity $v_{w}$ is
close to its initial value $v_i$ will give the main contribution to
$\Omega _{GW}$. Hence, we can use $v_i$ instead of $\bar{v}_w$ in the
approximations above. Inserting Eq. (\ref{eps}) in Eq. (\ref{om}) and
using $v_w\sim v_i$, $\dot{f}_{b}\sim Fv_{i}/d$, we have
\begin{equation}
\left. \Omega _{GW}\right\vert _{\mathrm{peak}}\sim 6\times
10^{-6}(\alpha v_i)^{8/3}F^{4/3}\left( \frac{d}{H_{\ast
}^{-1}}\right) ^{2}. \label{omapp}
\end{equation}
Again, we see that strong phase transitions  favor detection at LISA,
since the amplitude is maximized for large values of $d/H_{\ast
}^{-1}$ (which also give millihertz frequencies). Weak phase
transitions give smaller values of $d/H_{\ast }^{-1}$, and also
smaller values of $v_{i}$ and $\alpha $. Therefore, we will have very
low values of $\Omega _{GW}(f_{p})$, together with high values of the
frequency, $f_{p}>mHz$. In that case, the generated GWs will not be
detected by LISA, but may be detected by space-based interferometers
of second generation, such as BBO.

\begin{figure}[htb]
\centering \epsfysize=7cm
\leavevmode \epsfbox{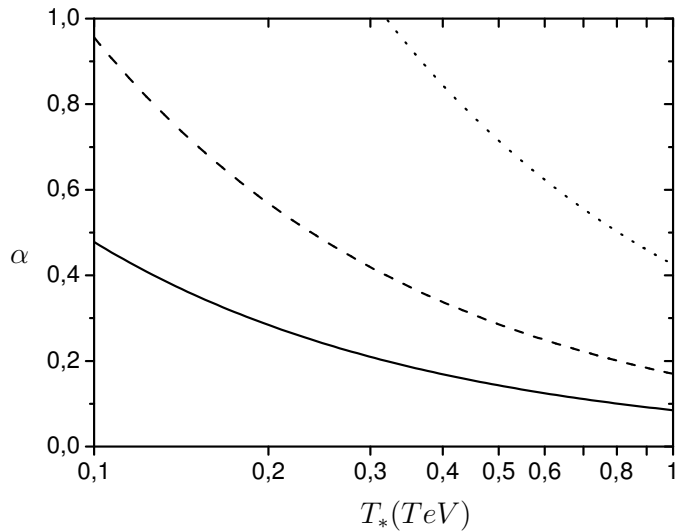}
\caption{The values of $\alpha$ and $T_*$ that give $f_p=1mHz$ and
$\Omega _{GW}(f_p)=10^{-11}$, for $v_i=0.1$
(solid line), $v_i=0.05$ (dashed line) and $v_i=0.02$ (dotted line).}
\label{figat}
\end{figure}

Let us consider an electroweak phase transition at  $T_{\ast
}\approx 100GeV$, with $g_{\ast }\approx 100$, and set $F\sim 10.$
%the characteristic frequency is $ f_{S}\sim 10^{-2}mHz\left( H_{\ast
%}^{-1}/d\right)$.
 The bubble separation
$d$ depends on the extension of the Standard Model. For the
frequency to be in the band that LISA is sensitive to, the phase
transition should be strongly first-order, so that $d/H_{\ast
}^{-1}\sim 10^{-2}$. Thus, for a wall velocity $v_{i}\sim 10^{-1}$
Eq. (\ref{omapp}) gives $\Omega _{GW}\sim 10^{-11},$ provided that
the latent heat is large enough (i.e., $\alpha \sim 1$). A large
latent heat is consistent with a strongly first-order phase
transition. This value of $\Omega _{GW}$ is just at the detection
threshold of LISA. Notice, however, that this is an
order-of-magnitude estimate, and the result is sensitive to several
parameters. For instance, the enthalpy difference $w_{+}-w_{-}$ in
Eq. (\ref{vfenth}) will give a larger efficiency factor if it is
evaluated at the supercooling temperature $T_{\Gamma }<T_{c}. $ For
a phase transition at $T_{\ast }=1TeV,$ Eq. (\ref{fs}) gives the
required frequency $f_{p}\sim mHz$ for a value $d/H_{\ast }^{-1}\sim
10^{-1} $. In this case, for $\alpha \sim 1$ and $v_i\sim 10^{-1}$
we have $\Omega _{GW}\sim 10^{-9}.$ In Fig. \ref{figat} we plot the
minimum value of $\alpha$ that is needed to achieve a peak value
$\Omega _{GW}(f_p)\geq 10^{-11}$ with a peak frequency $f_p=1mHz$
for a phase ransition that takes place at $T=T_*$.

\section{Conclusions\label{conclu}}

We have considered the production of gravitational waves in a
first-order phase transition, due to the turbulence that arises from
the motion of deflagration fronts. As we have seen, the dynamics of
the phase transition is completely different from the case of
detonations. The main difference is the fact that deflagration walls
are preceded by shock fronts, which move much faster than them and
influence other bubbles. This affects both the bubble expansion and
the generation of GWs. Firstly, as soon as bubbles begin to nucleate
and expand, the shock fronts collide and the spherical symmetry is
lost. Hence, gravitational radiation can be emitted before
percolation occurs. Secondly, the quick distribution of latent heat
causes a global reheating. As a consequence, the nucleation rate
turns-off after a very short time, and the bubble growth slows down.
We have taken these facts into account to calculate the spectrum of
GWs in a realistic scenario for deflagrations.

We have shown that the usual assumptions for the turbulent fluid,
namely, the validity of the Kolmogoroff spectrum and the short
duration of turbulence in comparison to the Hubble time, apply to the
case of deflagration bubbles. We have also derived an analytical
approximation for the efficiency factor $\kappa (\alpha,v_w)$ for the
deflagration case. In previous works this factor was either
calculated numerically or using rough estimations. This analytical
approximation makes the treatment of the deflagration case simpler,
and will be particularly helpful for including the GW calculation in
numerical computations of the phase transition.

We have used the results of Ref. \cite{cd06} for the GW spectrum from
primordial turbulence. There, it is assumed that the frequency
$\omega$ of a gravitational wave is determined by the wavenumber of
the source mode that generates it, i.e., $\omega=k$, rather than
being determined by the characteristic frequency of that mode,
$\omega_L=2\pi/\tau_{L}$. As a consequence, the GW spectrum inherits
the characteristic wavelength of the source, and the peak is at
$\omega_p\sim L_S^{-1}$. This is correct for a stochastic and
statistically homogeneous source of short duration. Recently
\cite{gkk07}, it was argued that the turbulent source lasts long
enough so that it can be treated as stationary. In that case, the
resulting GW spectrum would be imprinted with the characteristic
frequency of the source, and we would have $\omega_p\sim \omega_S$.
However, the duration of turbulence is on the order of the
characteristic turnover time $\tau_S\sim \omega_S^{-1}$ (since the
duration of the stirring source is shorter). Therefore, the
characteristic frequency and the duration of the source of GWs are
related by $\omega_S \tau_S={\cal O}(1)$, and the typical frequency
of the GWs is given by the typical wave number of the source
\cite{cds06}.

The GW spectrum depends on quantities such as  $v_{w},$
$\dot{f}_{b},$ and $d,$ which require a numerical computation of the
phase transition for a more accurate evaluation. We will address such
computation elsewhere \cite{progress}. Nevertheless, our estimations
show that GWs generated in a strongly first-order electroweak phase
transition might be detected by LISA. The parameter values for which
this  is possible are roughly constrained by $T_*\gtrsim 100GeV$,
$d\gtrsim 10^{-2}H_*^{-1}$, $v_w\gtrsim 10^{-2}$, $\alpha \gtrsim
0.1$. Weakly first-order phase transitions correspond in general to
smaller values of $d/H_*^{-1}$, $v_w$ and $\alpha$. In that case, the
spectrum will have a smaller amplitude and the characteristic
frequency will be away from the peak sensitivity of LISA.

\acknowledgements

I thank T. Kahniashvili for useful discussions. This work was
supported in part by Universidad Nacional de Mar del Plata,
Argentina, grant EXA 365/07, and by FONCyT grant PICT 33635.

\end{document}